\documentstyle[a4,11pt]{article}
\begin{document}

\begin{titlepage}
\vskip 2cm
\begin{flushright}
Preprint CNLP-1994-01
\end{flushright}
\vskip 2cm
\begin{center}
{\large {\bf ON MAGNETOELASTIC SOLITONS IN FERROMAGNET}}\footnote{Preprint
CNLP-1994-01.Alma-Ata.1994 }
\vskip 2cm

{\bf G.N.Nugmanova  }

\end{center}
\vskip 1cm
Centre for Nonlinear Problems, PO Box 30, 480035, Alma-Ata-35, Kazakhstan\\
E-mail: cnlpgnn@satsun.sci.kz

\vskip 1cm

\begin{abstract}
We study the solitonic excitations in the compressible
ferromagnetic Heisenberg chain (in the continuum limit).
\end{abstract}

%\maketitle

\end{titlepage}

\setcounter{page}{1}
\newpage
\begin{center}
{\bf INTRODUCTION}
\end{center}

Solitons in magnetically ordered crystals have been widely investigated
from both theoretical and experimental points of view[1-16]. In particular, the
existence of coupled magnetoelastic solitons in the Heisenberg compressible
spin chain has been extensively demonstrated[17-23]. In [23]
were presented the new class integrable and nonintegrable spin
systems. In this letter we consider the some of these  nonlinear models
of magnets - the some of the Myrzakulov equations(ME), which describe the
nonlinear dynamics of compressible magnets.
\\
\\
\begin{center}
A. {\bf THE 0-CLASS OF THE SPIN-PHONON SYSTEMS}
\end{center}

The Myrzakulov equations with the potentials have the form[23]:\\
the $ M^{10}_{00}$ - equation:
$$ 2iS_t=[S,S_{xx}]+(u+h)[S,\sigma_3] \eqno(1) $$
the $ M^{20}_{00}$ - equation:
$$ 2iS_t=[S,S_{xx}]+(uS_3+h)[S,\sigma_3] \eqno(2) $$
the $ M^{30}_{00}$ - equation:
$$ 2iS_t=\{(\mu \vec S^2_x-u+m)[S,S_x]\}_x+h[S,\sigma_3] \eqno(3) $$
the $ M^{40}_{00}$ - equation:
$$ 2iS_t=n[S,S_{xxxx}]+2\{(\mu \vec S^2_x-u+m)[S,S_x]\}_x+
h[S,\sigma_3] \eqno(4)$$
the $ M^{50}_{00}$ - equation:
$$ 2iS_t=[S,S_{xx}]+auS_x+bS_{x} \eqno(5) $$
where $v_{0}, \mu, \lambda, n, m, a, b, \alpha, \beta, \rho, h$ are constants,
$u$ is a scalar function(potential),
subscripts denote partial differentiations, $[,]$ (\{,\}) is
commutator (anticommutator),
$$S= \pmatrix{
S_3 & rS^- \cr
rS^+ & -S_3
}, \,\,\,\,\, S^{\pm}=S_{1}\pm i S_{2},\,\,\,\, r^{2}=\pm 1\,\,\,\,\, S^2=I.    $$

The solutions of these ME for the potential
$$ u=U sech^2 k(x-x_0)  \eqno(6a) $$
and for the boundary condition
$$
S\mid_{x=\pm\infty}=\sigma_3,\quad u\mid _{x=\pm\infty} =0 \eqno(7)
$$
are given by
$$
S^+ = AW shz \cdot sech^2 z, \quad S_3 = 1-2 sech^ 2 z \eqno(6b)
$$
and the following formulas, respectively $ (r^2=1, A^2=4,
W=exp(i(wt+\phi)), \phi=const, z=k(x-x_0)); $
$$ M^{30}_{00}: w=mk^2-h, k^2=U/4\mu, \lambda=\frac{1}{4} \eqno(8a)$$
$$ M^{40}_{00}: w=nk^4+2mk-h, k^2=U/(4\mu-5n) \eqno(8b)$$
\\
\\
\begin{center}
B.  {\bf THE 1-CLASS OF THE SPIN-PHONON SYSTEMS}
\end{center}

Here we present the following  ME[23]: \\
the $ M^{11}_{00}$ - equation:
$$
2iS_t=[S,S_{xx}]+(u+h)[S,\sigma_3] \eqno(9a)
$$
$$
\rho u_{tt}=\nu^2_0 u_{xx}+\lambda(S_3)_{xx} \eqno(9b)
$$
the $ M^{12}_{00}$ - equation:
$$
2iS_t=[S,S_{xx}]+(u+h)[S,\sigma_3] \eqno(10a)
$$
$$
\rho u_{tt}=\nu^2_0 u_{xx}+\alpha(u^2)_{xx}+\beta u_{xxxx}+
    \lambda(S_3)_{xx} \eqno(10b)
$$
the $ M^{13}_{00}$ - equation:
$$
2iS_t=[S,S_{xx}]+(u+h)[S,\sigma_3] \eqno(11a)
$$
$$
u_t+u_x+\lambda(S_3)_x=0 \eqno(11b)
$$
the $ M^{14}_{00}$ - equation:
$$
2iS_t=[S,S_{xx}]+(u+h)[S,\sigma_3] \eqno(12a)
$$
$$
u_t+u_x+\alpha(u^2)_x+\beta u_{xxx}+\lambda(S_3)_x=0 \eqno(12b)
$$
The some properties of these Myrzakulov equations were considered in refs.[25-28].
\\
\\
\begin{center}
C.  {\bf THE 2-CLASS OF THE SPIN-PHONON SYSTEMS}
\end{center}
In this section we consider the following ME[23]\\
the $ M^{21}_{00}$ - equation:
$$
2iS_t=[S,S_{xx}]+(uS_3+h)[S,\sigma_3] \eqno(13a)
$$
$$
\rho u_{tt}=\nu^2_0 u_{xx}+\lambda(S^2_3)_{xx} \eqno(13b)
$$
the $ M^{22}_{00}$ - equation:
$$
2iS_t=[S,S_{xx}]+(uS_3+h)[S,\sigma_3] \eqno(14a)
$$
$$
\rho u_{tt}=\nu^2_0 u_{xx}+\alpha(u^2)_{xx}+\beta u_{xxxx}+
\lambda (S^2_3)_{xx} \eqno(14b)
$$
the $ M^{23}_{00}$ - equation:
$$
2iS_t=[S,S_{xx}]+(uS_3+h)[S,\sigma_3] \eqno(15a)
$$
$$
u_t+u_x+\lambda(S^2_3)_x=0 \eqno(15b)
$$
the $ M^{24}_{00}$ - equation:
$$
2iS_t=[S,S_{xx}]+(uS_3+h)[S,\sigma_3] \eqno(16a)
$$
$$
u_t+u_x+\alpha(u^2)_x+\beta u_{xxx}+\lambda(S^2_3)_x=0 \eqno(16b)
$$
Some of these ME are studied in [25-28].
\\
\\
\begin{center}
D.  {\bf THE 3-CLASS OF THE SPIN-PHONON SYSTEMS}
\end{center}

Now we consider the following ME([23]):\\
the $ M^{31}_{00}$ - equation:
$$
2iS_t=\{(\mu \vec S^2_x - u +m)[S,S_x]\}_x  \eqno(17a)
$$
$$
\rho u _{tt}=\nu^2_0 u_{xx}+\lambda(\vec S^2_x)_{xx} \eqno(17b)
$$
the $M^{32}_{00}$ - equation:
$$
2iS_t=\{(\mu \vec S^2_x - u +m)[S,S_x]\}_x \eqno(18a)
$$
$$
\rho u _{tt}=\nu^2_0 u_{xx}+\alpha (u^2)_{xx}+\beta u_{xxxx}+ \lambda
(\vec S^2_x)_{xx} \eqno(18b)
$$
the  $M^{33}_{00}$ - equation:
$$
2iS_t=\{(\mu \vec S^2_x - u +m)[S,S_x]\}_x \eqno(19a)
$$
$$
u_t+u_x +\lambda (\vec S^2_x)_x = 0  \eqno(19b)
$$
the  $M^{34}_{00}$ - equation:
$$
2iS_t=\{(\mu \vec S^2_x - u +m)[S,S_x]\}_x \eqno(20a)
$$
$$
u_t+u_x +\alpha(u^2)_x+\beta u_{xxx}+\lambda (\vec S^2_x)_{x} = 0  \eqno(20b)
$$

The soliton solitons  of these ME - the $ M^{31}_{00},  M^{32}_{00}, M^{33}_{00}$
and $M^{34}_{00}$ equations are given by (6) and the following formulas, respectively
($r^2 = 1, A^2 =4, W=exp(i(\omega t +\varphi)),\omega=mk^4,\varphi = const,
z = k (x - x _0U = 4\mu k^2,$):
$$
M^{31}_{00}:\lambda = \mu\nu^2_0 \eqno(21a)
$$
$$
M^{32}_{00}:\alpha = 3\beta/2\mu, \quad k^2 = -(\lambda +\nu^2_0\mu)
/(4\mu\beta),\,\,\,\, \lambda=-\mu\nu^2_o-4\mu\beta k^2  \eqno(21b) $$
$$ M^{33}_{00}: \lambda=-\mu, k^2=U/4\mu \eqno(21c)$$
$$ M^{34}_{00}: \alpha=3\beta/2\mu, k^2=-(\lambda+\mu)/4\mu\beta \eqno(21d)$$
\\
\\
\begin{center}
E.  {\bf THE 4-CLASS OF THE SPIN-PHONON SYSTEMS}
\end{center}
These ME look like[23]:\\
the  $M^{41}_{00}$ - equation:
$$
2iS_t=[S,S_{xxxx}]+2\{((1+\mu)\vec S^2_x-u+m)[S,S_x]\}_{x} \eqno(22a)
$$
$$
\rho u_{tt}=\nu^2_0 u_{xx}+\lambda (\vec S^2_x)_{xx}  \eqno(22b)
$$
the  $M^{42}_{00}$ - equation:
$$
2iS_t=[S,S_{xxxx}]+2\{((1+\mu)\vec S^2_x-u+m)[S,S_x]\}_{x} \eqno(23a)
$$
$$
\rho u_{tt}=\nu^2_0 u_{xx}+\alpha(u^2)_{xx}+\beta u_{xxxx}+\lambda
(\vec S^2_x)_{xx} \eqno(23b)
$$
the  $M^{43}_{00}$ - equation:
$$
2iS_t=[S,S_{xxxx}]+2\{((1+\mu)\vec S^2_x-u+m)[S,S_x]\}_{x} \eqno(24a)
$$
$$
u_t + u_x + \lambda (\vec S^2_x)_x = 0   \eqno(24b)
$$
the  $M^{44}_{00}$ - equation:
$$
2iS_t=[S,S_{xxxx}]+2\{((1+\mu)\vec S^2_x-u+m)[S,S_x]\}_{x}\eqno{25a}
$$
$$
u_t + u_x + \alpha(u^2)_x + \beta u_{xxx}+\lambda (\vec S^2_x)_x = 0 \eqno(25b)
$$

These equations describe the nonlinear interaction of
the spin and phonon subsystems[23]. For the case $\mu=0$ the soliton solitons
of these equations were obtained in [7,28]. Here we consider the case $\mu \ne 0$.
In this case we present  the soliton solitons of the Myrzakulov
equatuons(22)-(25) for the boundary condition (7).
The soliton solitons of the ME  $M^{41}_{00},  M^{42}_{00},
M^{43}_{00} and M^{44}_{00}$ equations are given by (6) and the following formulas, respectively
($r^2 = 1, A^2 =4, W=exp(i(\omega t +\varphi)),\omega=k^4 +2mk^2,U = -yk^2,
y = 1 -4\mu, \varphi = const, z = k (x - x _0$)):
$$
M^{41}_{00}:\lambda = y\nu^2_0/4 \eqno(26a)
$$
$$
M^{42}_{00}:\alpha=-6\beta/ y, \quad k^2 = (4\lambda + \nu^2_0 y )/(4y\beta) \eqno(26b)
$$
$$
M^{43}_{00}:\lambda = y/4  \eqno(26c)
$$
$$
M^{44}_{00}:\alpha = -6\beta/y, \quad k^2 = (4\lambda - y )/(4y\beta) \eqno(26d)
$$
\\
\\
\begin{center}
F.  {\bf THE 5-CLASS OF THE SPIN-PHONON SYSTEMS}
\end{center}

Finally we consider the following equations[23]:\\
the  $M^{51}_{00}$ - equation:
$$ 2iS_t=[S,S_{xx}]+auS_x+bS_{x} \eqno(27a) $$
$$
\rho u_{tt}=\nu^2_0 u_{xx}+\lambda (f)_{xx}  \eqno(27b)
$$
the  $M^{52}_{00}$ - equation:
$$ 2iS_t=[S,S_{xx}]+auS_x+bS_{x} \eqno(28a) $$
$$
\rho u_{tt}=\nu^2_0 u_{xx}+\alpha(u^2)_{xx}+\beta u_{xxxx}+\lambda
(f)_{xx} \eqno(28b)
$$
the  $M^{53}_{00}$ - equation:
$$ 2iS_t=[S,S_{xx}]+auS_x+bS_{x} \eqno(29a) $$
$$
u_t + u_x + \lambda (f)_x = 0   \eqno(29b)
$$
the  $M^{54}_{00}$ - equation:
$$ 2iS_t=[S,S_{xx}]+auS_x+bS_{x} \eqno(30a) $$
$$
u_t + u_x + \alpha(u^2)_x + \beta u_{xxx}+\lambda (f)_x = 0 \eqno(30b)
$$
Here $f$ is a scalar function[23].
\\
\\
\begin{center}
Appendix. {\bf A LIST OF THE 2+1 DIMENSIONAL INTEGRABLE SPIN EQUATIONS}
\end{center}
Here we want present the some integrable (2+1)-dimensional spin systems -
the Ishimori and some Myrzakulov equations.

Consider a n-dimensional space with the basic unit vectors: $\vec e_{1}=
\vec S, \vec e_{2}, ... ,\vec e_{n}$ and $\vec e_{1}^{2}=E=\pm 1$. Then the 2+1 dimensional
Myrzakulov-0 equation[23]
has the form
$$
\vec S_{t} = \sum^{n}_{i=2} a_{i} \vec e_{i} \eqno(31a)
$$
$$
\vec S_{x} = \sum^{n}_{i=2} b_{i} \vec e_{i}           \eqno(31b)
$$
$$
\vec S_{y} = \sum^{n}_{i=2} c_{i} \vec e_{i} \eqno(31c)
$$
where $a_{i}, b_{i}, c_{i} $ are real functions, $\vec S = (S_{1}, S_{2}, ... , S_{n}),
\vec S^{2} = E = \pm 1$. This equation admits the many interesting class
integrable and nonintegrable reductions. Below we present only the
some integrable reductions of the Myrzakulov-0 equation.

1) The Myrzakulov-IV(M-IV) equation
$$ \vec S_{t}+\{\vec S_{xy}+V\vec S +
E\vec S_{x}\wedge (\vec S\wedge\vec S_{y})\}_{x}=0
$$
$$
V_{x}=\frac{E}{2} (\vec S^{2}_{x})_{y}
$$

2) The Myrzakulov-I(M-I) equation looks like
$$
\vec S_{t}=(\vec S\wedge \vec S_{y}+u\vec S)_x
$$
$$
u_x=-\vec S_(\vec S_{x}\wedge \vec S_{y})
$$

3) The Myrzakulov-II(M-II) equation
$$
\vec S_{t}=(\vec S\wedge \vec S_{y}+u\vec S)_x+2cb^{2}\vec S_{y}
     -4cv\vec S_{x}
$$
$$
u_x=-\vec S_(\vec S_{x}\wedge \vec S_{y}),
$$
$$ v_x=\frac{1}{16b^{2}c^{2}}
   (\vec S^2_{1x})_y
$$

4) The Myrzakulov-III(M-III) equation
$$ \vec S_{t}=(\vec S\wedge \vec S_{y}+u\vec S)_x+2b(cb+d)\vec S_{y}
     -4cv\vec S_{x}
$$
$$ u_x=-\vec S_(\vec S_{x}\wedge \vec S_{y}),
$$
$$ v_x=\frac{1}{4(2bc+d)^2}
   (\vec S^2_{1x})_y
$$
5) The Myrzakulov-VIII(M-VIII) equation looks like
$$ iS_t=\frac{1}{2}[S_{xx},S]+iuS_x
$$
$$
u_{y}=\frac{1}{4i}tr(S[S_y,S_x])
$$
where the subscripts denote partial derivatives and S denotes the spin
matrix $ (r^2=\pm1)$

$$S= \pmatrix{
S_3 & rS^- \cr
rS^+ & -S_3
},
$$
$$ S^2=I    $$

6) The Ishimori equation
$$ iS_t+\frac{1}{2}[S,M_{10}S]+A_{20}S_x+A_{10}S_y = 0 $$
$$ M_{20}u=\frac{\alpha}{4i}tr(S[S_y,S_x]) $$
where $ \alpha,b,a  $= consts and
$$ M_{j0} = M_{j},\,\,\, A_{j0}=A_{j}\,\,\,\, as \, \,\,\,a = b = -\frac{1}{2}. $$

7) The Myrzakulov-IX(M-IX) equation has the form
$$ iS_t+\frac{1}{2}[S,M_1S]+A_2S_x+A_1S_y = 0 $$
$$ M_2u=\frac{\alpha}{4i}tr(S[S_y,S_x]) $$
where $ \alpha,b,a  $=  consts and
$$ M_1= \alpha ^2\frac{\partial ^2}{\partial y^2}-2\alpha (b-a)\frac{\partial^2}
   {\partial x \partial y}+(a^2-2ab-b)\frac{\partial^2}{\partial x^2}; $$
$$ M_2=\alpha^2\frac{\partial^2}{\partial y^2} -\alpha(2a+1)\frac{\partial^2}
   {\partial x \partial y}+a(a+1)\frac{\partial^2}{\partial x^2},$$
$$ A_1=i\alpha\{(2ab+a+b)u_x-(2b+1)\alpha u_y\} $$
$$ A_2=i\{\alpha(2ab+a+b)u_y-(2a^2b+a^2+2ab+b)u_x\}, $$
The M-IX eqs. admit the two integrable reductions. As b=0,
after the some manipulations reduces to the M-VIII equation and as
 $ a=b=-\frac{1}{2} $
to the Ishimori equation. In general we have the two integrable cases:
 the M-IXA equation as $\alpha^{2} = 1,$ the M-IXB equation
as $\alpha^{2} = -1$. We note that the M-IX equation is integrable and
admits the following Lax representation
$$ \alpha \Phi_y =\frac{1}{2}[S+(2a+1)I]\Phi_x  $$
$$ \Phi_t=\frac{i}{2}[S+(2b+1)I]\Phi_{xx}+\frac{i}{2}W\Phi_x   $$
where $$ W_1=W-W_2=(2b+1)E+(2b-a+\frac{1}{2})SS_x+(2b+1)FS $$
$$ W_2=W-W_1=FI+\frac{1}{2}S_x+ES+\alpha SS_y $$
$$ E = -\frac{i}{2\alpha} u_x,\,\,\,  F = \frac{i}{2}(\frac{(2a+1)u_{x}}{\alpha} -
2u_{y}) $$

Hence  we get the Lax refresentations of the M-VIII
as $b = 0$ and for the Ishimori equation  as $a=b=-\frac{1}{2}$. The
M-IX equation  admit the different type exact solutions
(solitons, lumps, vortex-like, dromion-like and so on).

8) The Myrzakulov-XXII(M-XXII) equation has the form
$$ -iS_t=\frac{1}{2}([S,S_y]+2iuS)_x+\frac{i}{2}V_1S_x-2ia^2 S_y
$$
$$ u_x=-\vec S(\vec S_x\wedge \vec S_y) $$
$$ V_{1x}=\frac{1}{4a^2}(\vec S^2_x)_y $$
and so on.

All of these equations admit the corresponding Lax representations,
which were presented in [23]. The  gauge equivalent counterparts
of the above presented Myrzakulov equations are found in[28].

\end{document}